\documentstyle[prl,aps,epsf,multicol]{revtex}
\tighten
\begin{document}
\draft
\title{Quantum shot-noise at local tunneling contacts\\ on
mesoscopic multiprobe conductors}
\author{Thomas Gramespacher and Markus B\"uttiker}
\address{D\'epartement de Physique Th\'eorique, Universit\'e de Gen\`eve,
CH-1211 Gen\`eve 4, Switzerland}
\date{\today}
\maketitle
\begin{abstract}
New experiments that measure the low-frequency shot-noise spectrum at local
tunneling contacts on mesoscopic structures are proposed.  
The current fluctuation spectrum at a {\em single}
tunneling tip is determined by local partial densities of states. 
The current-correlation spectrum between {\em two} tunneling tips is 
sensitive to {\em non-diagonal} density of states elements
which are expressed in 
terms of products of scattering states of the conductor. 
Thus such an experiment permits to investigate correlations of electronic 
wave functions. We present specific results for a clean wire with 
a single barrier and for metallic diffusive conductors. 
\end{abstract}
\begin{multicols}{2}
\narrowtext
Since the original implementation of scanning tunneling microscopy
\cite{binnig82} a multitude of related scanning probe 
techniques\cite{wiesendanger,avouris95}
have permitted to obtain an unprecedented wealth of information
on the nanoscopic scale. It is the purpose of this work 
to present theoretical predictions of the shot noise measured at a point 
tunneling contact. Shot noise arises due to the quantization of the charge 
in the presence of transport \cite{BdJ}.  Measurements of the shot noise 
with a weak tunneling contact (such as the tip of an 
STM) are interesting not only because they would 
permit to create a map of the spatial distribution of the shot noise 
but also, as we will show, because they permit a measurement 
of the correlation of wave functions. This is in contrast to 
conductance or tunneling measurements which are related to density of 
states and thus to absolute squares of wave functions. 
Below we show that an investigation of the current-current correlation 
at two tunneling contacts permits to extract information also 
on the phase of an electronic wave function relative to that of another 
wave function. 

The typical arrangement in which scanning tunneling microscopy is 
used to investigate surface effects corresponds to a two 
terminal setup: the sample provides one terminal and the tip
provides the other terminal. In this case the tunneling current is proportional 
to the local density of states $\nu(x)$ at the location of the tip.
In this work we consider a mesoscopic structure that 
supports a transport current. Thus the sample must already have at least two
contacts which provide a source and sink for the carrier current (see Fig. 1).
In this case we have to treat a three-terminal structure, and it
depends in general on whether one 
is concerned with the tunneling conductance from the tip to the right
or left contact. Instead of the total density of states the 
tunneling conductance is related to a local partial density of states (LPDOS)
$\nu(x, \alpha)$ where $\alpha =1, 2$ labels the contacts of the conductor.
With the help of the density of states $1/hv_{\alpha m}$
of the $m$-th transverse scattering channel of reservoir $\alpha$ 
and the scattering states $\psi_{\alpha m}(x)$ incident 
from such a channel
the LPDOS can be expressed as 
\begin{equation}
\nu(x,\alpha)=\sum_{m\in \alpha}\frac{1}{hv_{\alpha m}}|\psi_{\alpha m}(x)|^2\, .
\label{lpdos}
\end{equation}
For a derivation of this result closely related to 
the discussion given below we refer to Ref. \cite{gramespacher97}. 
The LPDOS determines 
the charge injected from contact $\alpha$ into a region at position
$x$ in response 
to an increase of the Fermi energy of contact $\alpha$.
We can thus refer
to a LPDOS also as the injectivity of contact $\alpha$. 
The local density of states (LDOS) is  
the sum of the injectivities of all contacts,
$\nu(x)=\sum_\alpha\nu(x,\alpha)$. 
Below, as a first step, we show that the injectivities also determine 
the shot noise measured at a single tunneling contact 
(see Fig. 1, only tip 1 present). 
\begin{figure}[b]
\epsfxsize7cm
\centerline{\epsffile{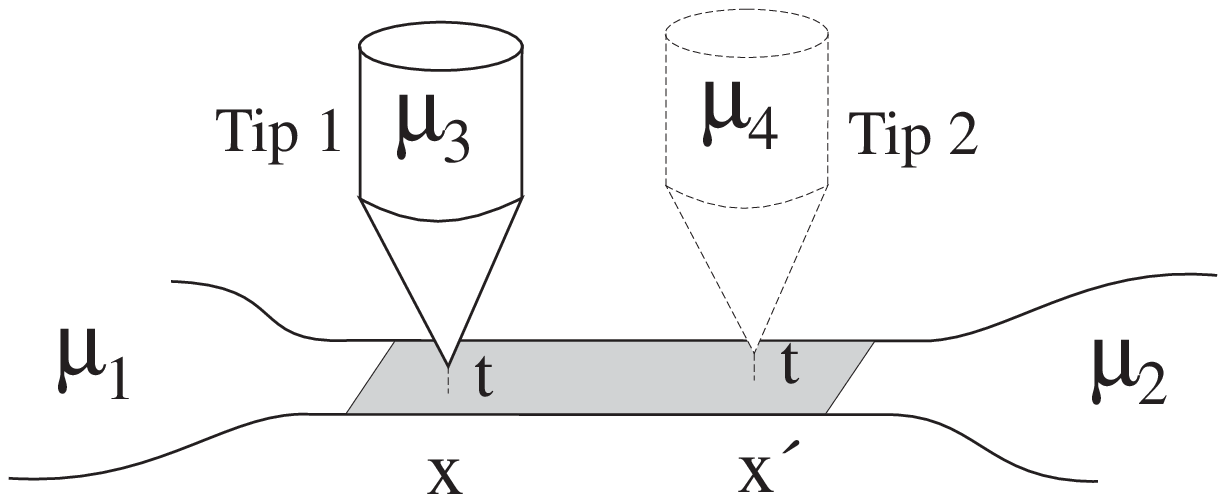}}
\vspace{0.3cm}
\caption{Mesoscopic conductor with contacts at potentials $\mu_1$
and $\mu_2$ and a tunneling contact at potential $\mu_3$, tip 1.
A second tunneling
contact tip 2 (dashed lines) at potential $\mu_4$ is only present for the
measurement
of the current-correlation spectrum.
The tunneling tips couple locally with strength $t$ at the points $x$ resp.\
$x^\prime$ to the wire.}
\end{figure}
In a second step, we consider two tunneling contacts, a four-terminal geometry,
and evaluate the correlation of the shot noise 
measured at these contacts. 
The correlation can not be expressed
with the help of the injectivities (which depend only on the absolute
square of wave functions) but are determined by non-diagonal
(non-local) elements of a density of states operator, 
\begin{equation}
\nu(x,x^\prime ,\alpha)=\sum_{m\in\alpha}\frac{1}{hv_{\alpha m}}\psi_{\alpha m}(x)
\psi_{\alpha m}^\star(x^\prime)\, .\label{injop}
\end{equation} 
We note that these elements are not real but depend on the 
phase difference which the wave function 
accumulates between the location of the two tips at $x$ and $x^\prime$.
The measurement of such a correlation permits thus the determination 
not only of the absolute square of the wave function but also of the phase
of the wave function. In a recent work 
Byers and Flatt\'{e} suggested a conductance experiment with two
tunneling probes on a surface\cite{byers95}.
They found that to second order in the tunneling
strength the current is determined by non-diagonal terms of the
Greens functions, i.\ e.\ spatial correlations of the wave functions.
In a conductance measurement spatial correlations represent
a small correction to a dominant first order term. 
In contrast, in the shot noise experiment proposed here, 
the wave function correlations provide the leading term. 
We illustrate our results for two particular geometries:
a ballistic wire which contains a single barrier and 
a metallic diffusive wire. 

There has been a continued strong interest in the shot noise 
of mesoscopic samples \cite{BdJ}. 
Since the initial experiments\cite{washburn91}
the development of highly sensitive and accurate 
measurement techniques \cite{reznikov95}
has permitted a close comparison between experimental techniques 
and theoretical
predictions \cite{BdJ,martin92,buttiker92}. 
It is thus justified to assume that 
similar techniques can be applied to the shot-noise measurement at tunneling 
contacts. 

Our theoretical starting point is a general formula 
which expresses the 
shot noise in mesoscopic multiprobe conductors
in terms of quadruples of scattering matrices\cite{buttiker92}. 
The spectrum of the current correlations in two contacts $\alpha$ and $\beta$
of a mesoscopic multiprobe conductor is defined as the Fourier
transform of the current-current correlator, $S_{\alpha\beta}(\omega)=\int dt
e^{i\omega t}\langle \Delta I_\alpha(t+t_0)\Delta I_\beta(t_0)\rangle$, where
$\Delta I_\alpha(t)=I_\alpha(t)-\langle I_\alpha(t)\rangle$ is the fluctuation
of the current in contact $\alpha$ away from its time-average.
In the low-frequency limit
the correlation spectrum can be expressed in terms of the current matrix
$A_{\delta\gamma}(\alpha)={\bf 1}_\alpha\delta_{\alpha\delta}
\delta_{\alpha\gamma}
-{\bf s}_{\alpha\delta}^\dagger(E) {\bf s}_{\alpha\gamma}(E)$
and the Fermi functions
$f_\delta(E)$ of the electron reservoirs\cite{buttiker92},
\begin{equation}
S_{\alpha\beta}=\frac{2e^2}{h}
\sum_{\delta\gamma}
\int dE Tr\left[ A_{\delta\gamma}(\alpha)A_{\gamma\delta}(\beta)\right]
f_\delta(1-f_\gamma)\, .\label{gennoise}
\end{equation}  
Here, ${\bf s}_{\alpha\beta}$ is the submatrix of the scattering matrix of
the sample which
describes scattering from all channels of contact $\beta$ into the channels
of contact $\alpha$.
We use Eq.\ (\ref{gennoise}) to find the fluctuation spectrum of the current
at the tunneling tip, $S_{33}$, as shown in Fig.\ 1 (only tip 1 present).
The tip couples locally at a point $x$ to the wire with a coupling strength
$t$.
We use the Hamiltonian formulation of the scattering matrix\cite{iida90}
to expand the
scattering matrix of the full system (wire and tip) to the lowest order
in the coupling strength $t$. The current
fluctuations in the tip can then be expressed with the help of the
scattering matrices of the two isolated systems and the coupling constant $t$.
We assume an applied voltage $eV=\mu_1-\mu_2$ at the two contacts
of the wire and set the electro-chemical potential at the tip 
$\mu_3=[\nu(x,1)\mu_1+\nu(x,2)\mu_2]/\nu(x)$ such that 
the average current into the tip
vanishes\cite{gramespacher97}.
With the two terminal tip to sample conductance ($\mu_1 = \mu_2$)
$G(x) = ({e^2}/{h}) 4\pi^2\nu_{tip}|t|^2 \nu(x)$,
where $\nu_{tip}$ is the LDOS of the isolated tip,
we find at zero temperature and in linear response to the applied 
potentials the shot-noise spectrum  
\begin{equation}
S_{33}= 2eG(x)V 
\frac{\nu(x,1)}{\nu(x)}\left( 1-\frac{\nu(x,1)}
{\nu(x)}\right)
\, . \label{fluct}
\end{equation}
Thus the noise is determined by 
$\nu(x,1)$, the injectivity of contact $1$ at the
coupling point $x$ in the wire, Eq.\ (\ref{lpdos}).
For small
potential differences all densities have to be taken at the Fermi energy.
Eq.\ (\ref{fluct}) suggests that the ratio $\nu(x,1)/\nu(x)$ 
plays the role of an effective local distribution function. It is
an exact quantum mechanical quantity which contains information
on the carrier propagation from contact $1$ all the way to the point
of observation. This is in contrast to the distribution functions used in the
semi-classical Boltzmann equation approach \cite{nagaev92,sukhorukov98,BdJ}
which contain no phase information.  
We now illustrate Eq.\ (\ref{fluct}) for the case of a perfect ballistic 
one channel conductor
with a barrier of transmission probability $T$ at $x=0$. 
At a position $x$ we find the injectivities 
\begin{eqnarray}
\nu(x,1) & = & \frac{1}{hv}(2-T+2\sqrt{1-T}\cos(2k_Fx))\, ,\\
\nu(x,2) & = & \frac{1}{hv}T\, .
\end{eqnarray}
As stated above the LDOS is the sum of the injectivities,
$\nu(x) = \nu(x,1)+\nu(x,2)$.
These densities together with the current fluctuations are shown in Fig.\ 2.
\begin{figure}
\epsfxsize8.5cm
\epsffile{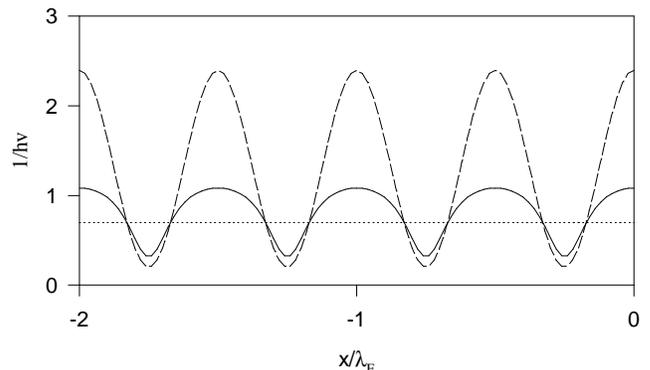}
\caption{Spatial variation of the current fluctuations and injectivities of
a ballistic wire with a barrier with transmission probability $T=0.7$.
The distance $x$ is measured relative to the barrier at $x=0$. 
The injectivity of the 
left contact $\nu(x,1)$, (dashed line), and the
injectivity of the right contact $\nu(x,2)$, (dotted line) are measured in units
of $1/hv$. The solid line is the current fluctuation spectrum $S_{33}$,
Eq.\ (\ref{ball}).}
\end{figure}
As a function of the tip position $x$, the fluctuation spectrum
\begin{eqnarray}
S_{33} \propto T\left( 1-\frac{T}{2}\frac{1}{1+\sqrt{1-T}\cos(2k_Fx)}\right)
\label{ball}
\end{eqnarray}
shows an oscillating
behavior with the period of half a Fermi wavelength 
$\lambda_F = 2\pi/k_F$.
If we average this spectrum over one oscillation period, we find
$\langle S_{33}\rangle_{ave} \propto
T(1-\sqrt{T}/2)$. Note that this differs from the fluctuation spectrum 
that would be measured at a massive contact, $S_{11} \propto T(1-T)$.
The dependence on $\sqrt{T}$ instead of $T$ has its origin
in the interference of incident and reflected waves. 
It is tempting to say that the fluctuations
in the tip reflect directly the intrinsic fluctuations in the wire.
Note, however,
that even though a perfect
ballistic wire ($T=1$)
shows no shot noise, the current in a tip which probes such
a wire would fluctuate. For $T =1$, the right hand side of Eq.\ (\ref{ball})
does not vanish but is 
$1/2$.

As a second example, we investigate a metallic diffusive wire. 
The diffusive wire
extends from $x=0$ to $x=L$, and has a width $W$ much smaller
than its length $L$. For the ensemble averaged quantities, 
the diffusion can then be considered to be one-dimensional.
Furthermore we assume that $k_Fl\gg1$ with the elastic mean free path $l\ll L$.
The ensemble averaged injectivities
of the two contacts of the wire are in the diffusive region\cite{christen}
$\nu(x,1)=\nu_0\frac{L-x}{L}$ and 
$\nu(x,2)=\nu_0\frac{x}{L}$ 
where $\nu_0=m^\star/(h\hbar)$ is the
two-dimensional density of
states and $m^\star$ the effective electron mass. 
In particular, the injectivities are independent of the transverse
coordinate. Using these densities in
Eq.\ (\ref{fluct}) gives a parabolic dependence of the
fluctuation spectrum on the tip position,
\begin{equation}
S_{33} \propto x(L-x)/L^2\, .
\end{equation}
Note that if we average this spectrum over the entire wire (from $x =0$ to 
$x = L$) this leads to a noise spectrum which is $1/3$ of that 
measured at a tunneling contact of a perfect ballistic wire. 
Again, we have the surprising similarity to the well known 1/3 reduction of
the shot noise at an isolated metallic diffusive
conductor\cite{beenakker92,nagaev92,steinbach96}.

Next we investigate the spectrum $S_{34}$ of the
correlations of the currents in two tips which couple at positions $x$
and $x^\prime$ to a wire (Fig.\ 1, tip 1 and tip 2 present).
Again, we can start from the general
formula, Eq.\ (\ref{gennoise}), and expand the scattering matrix of the entire
system
to the lowest order in the coupling strength $t$. The result for the correlation
spectrum depends in general on the electro-chemical potentials at all
four contacts of the system.
Here, we specialize to three different configurations
of the applied voltages. We call these configurations experiment A,B and C.
First, all potentials are held at the equilibrium value $\mu_0$. In experiment
A now, we rise the potential of the left contact of the wire (contact 1) to
the elevated value $\mu$, so that current is injected into the system through
this contact. In experiment B we rise only the potential of the right contact
of the wire (contact 2) to the value $\mu$ all others being held at the
equilibrium potential $\mu_0$. In experiment C we rise simultaneously
the potentials of both sides of the wire (contacts 1 and 2) to the value $\mu$.
Comparison of the correlations of experiments A,B and C permits 
to identify the exchange correlations, i.\ e.\ the effect due to the 
quantum mechanical indistinguishability of particles\cite{buttiker92}. 
Exchange effects in metallic diffusive conductors with wide contacts
are the subject of Refs.\ \cite{sukhorukov98} and \cite{blanter97}.
Ballistic cavities with four tunneling contacts are investigated 
in Ref.\ \cite{langen97}. An experiment by Liu {\em et al.\ }\cite{liu98}
measures exchange effects in an open ballistic structure. 
At $kT=0$ and in linear response to the applied bias
$eV=\mu-\mu_0$, we find for the correlation spectrum at the 
two tunneling tips,
\begin{equation}
S_{34} = 2\frac{e^2}{h}
eV16\pi^4\nu_{tip1}\nu_{tip2}|t|^4S_{A,B,C}
\end{equation}
with $\nu_{tip\alpha}$ being the LDOS in tip $\alpha$ and
\begin{eqnarray}
S_A & = & -2|\sum_{m\in 1}\frac{1}{hv_{1m}}\psi_{1m}(x)
\psi_{1m}^\star(x^\prime)|^2\, ,\label{expa}\\
S_B & = & -2|\sum_{n\in 2}\frac{1}{hv_{2n}}\psi_{2n}(x)
\psi_{2n}^\star(x^\prime)|^2\, ,\label{expb}\\
S_C & = & -2|\sum_{\alpha=1,2}\sum_{m\in\alpha}\frac{1}
{hv_{\alpha m}}\psi_{\alpha m}(x)\psi_{\alpha m}^\star(x^\prime)|^2\nonumber \\
& = & S_A+S_B-4\sum_{{m\in 1}\atop{n\in 2}}\frac{1}
{h^2v_{1m}v_{2n}}\nonumber \\
& & \times 
Re\{ \psi_{1m}(x)\psi_{2n}^\star(x)\psi_{1m}^\star(x^\prime)
\psi_{2n}(x^\prime) \} \, .\label{expc}
\end{eqnarray}
Here, $\psi_{\alpha m}(x)$ is the scattering state describing an incoming
electron in channel $m$ of contact $\alpha$ which is scattered into all
channels of both contacts of the wire. The velocities $v_{\alpha m}
=\sqrt{2(E_F-E_{\alpha m}^0)/m^\star}$ with $E_{\alpha m}^0$ being the threshold
energy of channel $m$ in contact $\alpha$ and $m^\star$ being the electron
mass are evaluated at the Fermi energy $E_F$.
The sums are over all open channels
in contact 1 resp.\ contact 2.

To arrive at these results which express 
the noise correlations in terms of 
scattering states, we proceed as
follows: we express the scattering matrix in Eq.\ \ref{gennoise} in terms of
the Greens function of the four-probe sample (wire and tips) and the coupling
matrix which
couples the ideal leads to the mesoscopic sample. We expand the Greens
function to first order in the weak links $t$ between the tips and the wire. The
scattering states are finally related to the Greens function of the sample
and the coupling matrix between the leads and
the sample by a Lippmann-Schwinger equation. 
Note that with the help of the injectivity operator Eq.\ \ref{injop},
we can express Eqs.\ (\ref{expa})-(\ref{expc}) in the following compact form:
$S_A= - 2 |\nu(x,x^\prime,1)|^2$, $S_B
=-2 |\nu(x,x^\prime,2)|^2$, and $S_C=-2|\nu(x,x^\prime,1)+
\nu(x,x^\prime,2)|^2$.

We now use the results, Eqs.\ (\ref{expa})-(\ref{expc}),
to investigate the current
correlations for the diffusive wire discussed above. 
We are interested in the correlations averaged over impurity configurations. 
For the averaging procedure we assume that the distance between the two tips
and between each tip and the boundaries of the diffusive region is much larger
than the elastic mean free path $l$. We express the wave functions
in terms of four Greens functions and use the diagram technique to average
the products of Greens functions\cite{altshuler85}.
It turns out, that for all three experiments, the
strongest contribution to the averaged quantity comes from diagrams 
which contain four diffusons\cite{blanter97}.
Diagrams with two and three diffusons are small as $l/L$
resp.\ $(l/L)^2$. With the abbreviation
$a(x,x^\prime)=1/3[(x-x^\prime)^2-2x^\prime(L-x)]$
the leading order terms are,
\begin{eqnarray}
S_A & = & \frac{S_C}{2}
\frac{(L-x)^2+(L-x^\prime)^2+a(x,x^\prime)}{L^2}\, ,\\
S_B & = & \frac{S_C}{2}
\frac{x^2+{x^\prime}^2+a(x,x^\prime)}{L^2}\, ,\\
S_C & = & -2
\frac{4(m^\star)^2}{(h\hbar)^2N}
\frac{L}{l}\frac{x(L-x^\prime)}{L^2}
= -4\frac{\nu(x,2)\nu(x^\prime,1)}{g}\, .
\end{eqnarray}
where $g=\frac{l}{2L}N$ is the Drude conductance and $N=k_FW$ is the number of
channels. 
Here, we assumed that tip 1 is positioned to the left of tip 2.
At once we see that even after averaging over the impurity configurations, the result
of experiment C is not just the addition of experiments A and B. In fact,
it is interesting to determine the strength of the exchange term $S_X=
S_C-S_A-S_B$. In general, this expression depends on the two coordinates
$x$ and $x^\prime$. 
We investigate it closer for the special case where the tips are placed
symmetrically around the center of the wire, $L/2$, i.\ e.\ tip 1 is placed at
a distance
$d/2$ to the left of the center and tip 2 is placed at the same distance $d/2$
to the right of the center. The relative strength of the exchange term is then
as a function of the distance $d$ between the tips
\begin{equation}
\frac{S_X}{S_C}=\frac{1}{3}\left(2+\frac{d}{L}-2\left(\frac{d}{L}\right)^2
\right)\, .
\end{equation}
Interestingly, this function reaches its maximum not when the tips are closest, but
at the finite distance $d=L/4$ (which is still large compared to $l$).
Its maximal value is $(S_X/S_C)_{max}=17/24$. For any two tip positions $x$
and $x^\prime$, the exchange term $S_X$ is always negative and therefore
enhances the current correlations. An enhancement of the current correlations by
the exchange term was also predicted for a chaotic cavity with four
tunneling contacts\cite{langen97}.

In conclusion, we have shown that noise measurements at local tunnel junctions
can reveal considerably more information about
the electronic structure of a mesoscopic system than are accessible to pure
conductance measurements. In the case of a single tip the shot noise
is determined by an effective local distribution function, 
$\frac{\nu(x,\alpha)}{\nu(x)}$. 
Especially interesting are the current correlation spectra of two tips.
For the three suggested experiments, Eqs.\ (\ref{expa})-(\ref{expc}), they
depend directly on the phase-carrying amplitudes of the wave functions. 
These experiments can be used to demonstrate the importance of the
exchange-correlation due to the indistinguishability of the electrons. 
We have shown that for a metallic diffusive wire, the exchange term gives
always a negative contribution to the correlation spectrum 
(enhances the effect) which can be as high as 70\% of 
the total correlation spectrum. 

We thank Ya.\ M.\ Blanter for his advice on the
use of the diagram technique for metallic diffusive conductors.

This work was supported by the Swiss National Science Foundation.
\vspace{-0.6cm}
\end{multicols}

\begin{references}
%
\vspace{-1.8cm}
%
\bibitem{binnig82}
G.\ Binnig and H.\ Rohrer, Helv.\ Phys.\ Acta {\bf 55}, 726 (1982);
G.\ Binnig {\em et al.}, Phys.\ Rev.\ Lett.\ {\bf 49}, 57 (1982).
%
\bibitem{wiesendanger}
{\em Scanning Tunneling Microscopy I,II,III}, Springer Series in Surface Sciences 20,
28, 29, Ed.\ by R.\ Wiesendanger and H.-J.\ G\"untherodt (Springer,
Heidelberg, 1992).
%
\bibitem{avouris95}
Ph.\ Avouris, I.-W.\ Lyo, and Y.\ Hasegawa, IBM J.\ Res.\ Dev.\ {\bf 39},
603 (1995), and other articles in the same issue.
%
\bibitem{BdJ} M.~J.~M.~de~Jong and C.~W.~J.~Beenakker, in:
{\em Mesoscopic Electron Transport}, Ed.\ by L.~L.~Sohn,
L.\ P.\ Kouwenhoven, and G.\ Sch\"on, NATO
ASI Series E, Vol. {\bf 345} (Kluwer, Dordrecht, 1997),
p.\ 225.
%
\bibitem{gramespacher97}
T.\ Gramespacher and M.\ B\"uttiker, Phys.\ Rev.\ B {\bf 56}, 13026 (1997).
%
\bibitem{byers95}
J.\ E.\ Byers and M.\ E.\ Flatt\'{e}, Phys.\ Rev.\ Lett.\ {\bf 74}, 306 (1995).
%
\bibitem{washburn91}
Y.\ P.\ Li {\em et al.}, Appl.\ Phys.\ Lett.\ {\bf 57}, 774 (1990);
S.\ Washburn {\em et al.}, Phys.\ Rev.\ B {\bf 44}, 3875 (1991).
%
\bibitem{reznikov95} 
H.\ Birk, M.\ J.\ M.\ de Jong, and C.\ Schoenenberger, Phys.\ Rev.\ Lett.\ 
{\bf 75}, 1610 (1995);
M.\ Reznikov {\em et al.}, {\em ibid.} {\bf 75}, 3340 (1995);
A.\ Kumar {\em et al.}, {\em ibid.} {\bf 76}, 2778 (1996);
R.~J.\ Schoelkopf {\em et al.},
{\em ibid.} {\bf 78}, 3370 (1997).
%
\bibitem{martin92}
Th.\ Martin and R.\ Landauer, Phys.\ Rev.\ B {\bf 45},
1742 (1992).
%
\bibitem{buttiker92}
M.\ B\"uttiker, Phys.\ Rev.\ B {\bf 46}, 12485 (1992);
Phys.\ Rev.\ Lett.\ {\bf 68}, 843 (1992).
%
\bibitem{iida90}
S.\ Iida, H.\ A.\ Weidenm\"uller, and J.\ Zuk, Phys.\ Rev.\ Lett.\ {\bf 64},
583 (1990); Ann.\ Phys.\ (N.Y.) {\bf 200}, 219 (1990).
%
\bibitem{nagaev92} K.~E.~Nagaev, Phys.\ Lett.\ A {\bf 169}, 103 (1992).
%
\bibitem{sukhorukov98}
E.\ V.\ Sukhorukov and D.\ Loss, cond-mat/9802050
%
\bibitem{christen} M.\ B\"uttiker and T.\ Christen, in:
{\it Quantum Transport in Semiconductor Submicron Structures},
Ed.\ by B.\ Kramer, NATO ASI Series, Vol.\ {\bf 326} (Kluwer,
Dordrecht, 1996), p.\ 263.
%
\bibitem{beenakker92}
C.\ W.\ J.\ Beenakker and M.\ B\"uttiker, Phys.\ Rev.\ B {\bf 46}, 1889 (1992).
%
\bibitem{steinbach96}
A.\ H.\ Steinbach, J.\ M.\ Martinis, and M.\ H.\ Devoret,
Phys.\ Rev.\ Lett.\ {\bf 76}, 3806 (1996).
%
\bibitem{blanter97}
Ya.\ M.\ Blanter and M.\ B\"uttiker, Phys.\ Rev.\ B {\bf 56}, 2127 (1997).
%
\bibitem{langen97}
S.\ A.\ van Langen and M.\ B\"uttiker, Phys.\ Rev.\ B {\bf 56}, R1680 (1997).
%
\bibitem{liu98} R. Liu {\em et al.}, Nature {\bf 391}, 263 (1998). 
%
\bibitem{altshuler85}
B.\ L.\ Altshuler and A.\ G.\ Aronov, in: {\em Electron-electron Interactions
in Disordered Systems}, Ed.\ by A.\ L.\ Efros and M.\ Pollak (North-Holland,
Amsterdam, 1985), p.1.
%
\end{references}
\end{document}